\documentclass[a4paper]{article}

\usepackage[margin=1in]{geometry}

\usepackage[T1]{fontenc}
\newcommand{\changefont}[3]{
\fontfamily{#1} \fontseries{#2} \fontshape{#3} \selectfont}

\changefont{ptm}{m}{n}

\usepackage{setspace} 
\usepackage{graphicx}    

\usepackage{amsfonts}
\usepackage{amssymb}
\usepackage{graphics}
\usepackage{mathrsfs}
\usepackage{color}

\newtheorem{theorem}{Theorem}[section]

\newtheorem{corollary}{Corollary}[section]
\newtheorem{lemma}{Lemma}[section]

\newtheorem{definition}{Definition}[section]

\long\def\symbolfootnote[#1]#2{\begingroup%
\def\thefootnote{\fnsymbol{footnote}}\footnote[#1]{#2}\endgroup} 

\begin{document}

\begin{center}
\Large \textbf{Almost Periodicity in Chaos}
\end{center}

\begin{center}
\normalsize \textbf{Marat Akhmet$^{a,}\symbolfootnote[1]{Corresponding Author Tel.: +90 312 210 5355,  Fax: +90 312 210 2972, E-mail: marat@metu.edu.tr}$ and Mehmet Onur Fen$^b$} \\
\vspace{0.2cm}
\textit{\textbf{\footnotesize$^a$Department of Mathematics, Middle East Technical University, 06800 Ankara, Turkey \\ \footnotesize$^b$Basic Sciences Unit, TED University, 06420 Ankara, Turkey}}
\vspace{0.1cm}
\end{center}

\vspace{0.3cm}

\begin{center}
\textbf{Abstract}
\end{center}

\noindent\ignorespaces

Periodicity plays a significant role in the chaos theory from the beginning since the skeleton of chaos can consist of infinitely many unstable periodic motions. This is true for chaos in the sense of Devaney \cite{Dev90}, Li-Yorke \cite{Li75} and the one obtained through period-doubling cascade \cite{Feigenbaum80}. Countable number of periodic orbits exist in any neighborhood of a structurally stable Poincar\'{e} homoclinic orbit, which can be considered as a criterion for the presence of complex dynamics \cite{Gonchenko96}-\cite{Smale65}. It was certified by Shilnikov \cite{Shilnikov02} and Seifert \cite{Seifert97} that it is possible to replace periodic solutions by Poisson stable or almost periodic motions in a chaotic attractor. Despite the fact that the idea of replacing periodic solutions by other types of regular motions is attractive, very few results have been obtained on the subject. The present study contributes to the chaos theory in that direction. 

In this paper, we take into account chaos both through a cascade of almost periodic solutions and in the sense of Li-Yorke such that the original Li-Yorke definition is modified by replacing infinitely many periodic motions with almost periodic ones, which are separated from the motions of the scrambled set. The theoretical results are valid for systems with arbitrary high dimensions. Formation of the chaos is exemplified by means of unidirectionally coupled Duffing oscillators. The controllability of the extended chaos is demonstrated numerically by means of the Ott-Grebogi-Yorke \cite{Ott90} control technique. In particular, the stabilization of tori is illustrated.

\vspace{0.2cm}
 
\noindent\ignorespaces \textbf{Keywords:} Chaos with almost periodic motions; Li-Yorke chaos; Cascade of almost periodic solutions; Control of chaos; Stabilization of tori

\vspace{0.6cm}


\section{Introduction} \label{almost_periodicity_intro}

The transition between regular and chaotic motions has been a field of interest among scientists since the second half of the twentieth century. Investigations of chaos for continuous-time dynamics started with the studies of Poincar\'{e} \cite{Andersson94}, Cartwright and Littlewood \cite{Cartwright1}, Levinson \cite{Levinson}, Lorenz \cite{Lorenz63}, and Ueda \cite{Ueda78}. Chaos theory has many applications in various disciplines such as neuroscience, economics, mechanics, electronics, meteorology, and medicine \cite{Grebogi97}.

The first mathematical definition of chaos was introduced in the paper \cite{Li75} for discrete-time systems. According to Li and Yorke \cite{Li75}, a continuous map on an interval has points of all periods provided that it has a point of period three, and there exists an uncountable scrambled subset \cite{Huang02} of the interval. The part of the Li-Yorke Theorem concerning periodic motions is a specific version of the Sharkovskii Theorem \cite{Sharkovskii}. The presence of the scrambled set is the feature that distinguishes Li-Yorke chaos from other types \cite{Huang02}. Besides, a possible way for systems to achieve chaos is the period-doubling cascade, which was first observed in quadratic maps by Myrberg \cite{Myrberg58}-\cite{Myrberg63}. In this phenomenon, as some experimental parameter of the considered system varies, a motion with a fundamental period changes to a periodic motion with twice the period of the original oscillation, and as the parameter is changed further, the same procedure occurs. The period-doubling bifurcation values of the parameter accumulates at a finite value, after which chaos is observable \cite{Feigenbaum80,Sander11,Sander12}. The period-doubling onset of chaos exhibits universal behavior \cite{Feigenbaum80}, and it can be observed in many nonlinear systems, particularly in neural networks \cite{Barrio14}, ecological models \cite{Lacitignola10}, nonlinear circuits \cite{Hanias09}, and semiconductor lasers \cite{Simpson94}.

Throughout the paper, $\mathbb R,$ $\mathbb N$ and $\mathbb Z$ will denote the sets of real numbers, natural numbers and integers, respectively. 
The main object of the present investigation is the dynamics of the following system,
\begin{eqnarray} \label{new_replicator}
y'=Ay+G(t,y)+H(x(t)),
\end{eqnarray}
where the function $G:\mathbb R \times \mathbb R^n \to \mathbb R^n$ is continuous in both of its arguments and it is almost periodic in $t$ uniformly for $y\in \mathbb R^n,$ the function $H:\mathbb R^m \to \mathbb R^n$ is continuous, and all eigenvalues of the constant $n \times n$ real valued matrix $A$ have negative real parts.
Equation (\ref{new_replicator}) is obtained by perturbing the system 
\begin{eqnarray} \label{new_base_system}
u'=Au+G(t,u)
\end{eqnarray}
with the term $H(x(t)).$ The perturbation function $x(t)$ will be explained in a detailed form in Section \ref{ap_cascade_section} and Section \ref{ap_Li-Yorke_Section}. Our purpose is to prove rigorously that if the perturbation is chaotic, then system (\ref{new_replicator}) possesses chaos with \textit{infinitely many almost periodic motions}. We will make use of coupled Duffing oscillators to illustrate the formation of the chaos. Moreover, the controllability of the generated chaos will be demonstrated numerically. 

In the present study, two different types of chaos formation will be presented. In the first one, solutions of a system of differential equations that possesses a period-doubling cascade will be utilized as the perturbation function $x(t)$ in (\ref{new_replicator}). Secondly, we will make use of chaotic sets of functions as the source for the perturbation $x(t)$ to obtain Li-Yorke chaos with infinitely many almost periodic motions, which are separated from the scrambled set.

The main difficulty of indicating almost periodicity in chaos lies in the fact that systems which generate chaos do this with periodical scenario. Being based on special input-output mechanisms, the chaos replication method introduced and developed in our studies \cite{Akh2}-\cite{Akh22} allows to superpose solutions with incommensurate periods and to apply chaotic perturbation to the system with purely almost periodic motion to create infinitely many unstable almost periodic motions embedded in the chaotic attractor. 

The paper \cite{Akh4} deals with the general technique of dynamical synthesis, which was developed in \cite{Brown93}-\cite{Brown01}. The study \cite{Akh8} was concerned with the extension of chaos in the sense of Devaney \cite{Dev90}, Li-Yorke \cite{Li75} and the one obtained through period-doubling cascade \cite{Feigenbaum80,Sander11,Sander12} for unidirectionally coupled systems. The appearance of infinitely many quasi-periodic motions in a chaotic attractor was revealed in \cite{Akh8}. For that purpose, a system with an asymptotically stable equilibrium point was influenced by two chaotic systems that possess infinitely many periodic motions with incommensurate periods giving rise to quasi-periodicity. However, in the present study, we make use of only a single source of perturbations to obtain chaos with infinitely many \textit{almost periodic motions}, and the perturbed system has a different structure.

The idea that unstable periodic motions embedded in a chaotic attractor can be replaced by more general types of regular motions stems from the investigations of Poincar\'{e} and Birkhoff \cite{Nemytskii60}. In any neighborhood of a structurally stable Poincar\'{e} homoclinic orbit, there exist nontrivial hyperbolic sets containing a countable number of saddle periodic orbits and continuum of non-periodic Poisson stable orbits \cite{Gonchenko96}-\cite{Smale65}. According to Seifert \cite{Seifert97}, there exists discrete chaos with infinitely many almost periodic motions. It was certified in the paper \cite{Shilnikov02} that, in general, in place of a countable set of periodic solutions to form chaos, one can take an uncountable collection of Poisson stable motions which are dense in a quasiminimal set. This can also be observed in the Horseshoe attractor \cite{Smale67}. Motivated by these studies, continuous chaotic attractors with infinitely many almost periodic motions are considered in the present paper.

The concept of snap-back repellers for high dimensional maps was introduced in \cite{Marotto78}. According to Marotto \cite{Marotto78}, if a multidimensional continuously differentiable map has a snap-back repeller, then it is Li-Yorke chaotic. Marotto's Theorem was used in \cite{PLi07} to prove the existence of Li-Yorke chaos in a spatiotemporal chaotic system. Li-Yorke sensitivity, which links the Li-Yorke chaos with the notion of sensitivity, was studied in \cite{Akin03}. Moreover, generalizations of Li-Yorke chaos to mappings in Banach spaces and complete metric spaces were provided in \cite{Kloeden06}-\cite{Shi05}. In the present paper, we develop the concept of Li-Yorke chaos with infinitely many almost periodic motions, which are separated from the motions of a scrambled set, and provide a method for its formation.

Almost periodic and in particular quasi-periodic motions have an important place in the theory of neural networks. In the book \cite{Izhikevich07}, the dynamics of the brain activity is considered as a system of many coupled oscillators with different incommensurable periods. According to Pasemann et al. \cite{Pasemann2003}, quasi-periodic solutions are noteworthy in biological and artificial systems since they are associated with various kinds of central pattern generators. Watanabe et al. \cite{Watanabe97} demonstrated that chaotic dynamics works as means to learn new patterns and increases the \textit{memory capacity} of neural networks. The consideration of infinitely many almost periodic motions instead of periodic ones provides dynamics with a \textit{higher complexity}. Therefore, our results may be useful for neural networks to obtain a memory with a \textit{larger capacity} than that with periodic motions.

The extension of chaos through unidirectional couplings has been considered in the theory of synchronization \cite{Gon04}-\cite{Afraimovich01}. In our study, we do not consider the chaos synchronization problem, but we say that chaos with infinitely many almost periodic motions is generated. The asymptotic proximity of the solutions underlies the presence of synchronization. However, we do not take into account the solutions from the asymptotic point of view. Furthermore, in the synchronization of chaotic systems, one of the assumptions is that the response system is chaotic in the absence of the driving. On the contrary, in our case, the unperturbed system (\ref{new_base_system}) possesses an asymptotically stable almost periodic solution, and therefore, it is non-chaotic.

The remaining parts of the paper is organized as follows. In Section \ref{almost_periodicity_prelim}, we investigate the bounded solutions of (\ref{new_replicator}) and their attractiveness property. Section \ref{ap_cascade_section} is devoted for chaos through a cascade of almost periodic solutions. In Section \ref{ap_Li-Yorke_Section}, we describe the Li-Yorke chaos with infinitely many almost periodic motions and deal with the theoretical results concerned with its appearance in system (\ref{new_replicator}).  An example is provided in Section \ref{almost_periodicity_example} by means of unidirectionally coupled Duffing oscillators. In Section \ref{ap_control_section}, the control of the extended chaos as well as the stabilization of tori embedded in the chaotic attractor of the coupled Duffing oscillators are numerically demonstrated. Finally, some concluding remarks are given in Section \ref{ap_conclusion}.

\section{Preliminaries}\label{almost_periodicity_prelim}

In the remaining parts of the paper, we will make use of the usual Euclidean norm for vectors and the norm induced by the Euclidean norm for square matrices.

Since the matrix $A$ in system (\ref{new_replicator}) has eigenvalues all with negative real parts, one can verify that there exist positive numbers $N$ and $\omega$ such that $\left\|e^{At}\right\| \leq Ne^{-\omega t}$ for all $t\geq 0.$ 

The following conditions are required.

\begin{enumerate}
\item[\bf (C1)] There exists a positive number $L_1< \displaystyle \frac{\omega}{N}$ such that $\left\|G(t,y_1)-G(t,y_2)\right\| \leq L_1\left\|y_1-y_2\right\|$ for all $t\in \mathbb R,$ $y_1,$ $y_2 \in \mathbb R^{n};$
\item[\bf (C2)] There exists a positive number $M_G$ such that $\displaystyle \sup_{t\in \mathbb R, ~y\in \mathbb R^n} \left\|G(t,y)\right\|\le M_G.$
\end{enumerate}

Utilizing the theory of quasilinear equations \cite{Hale80}, it can be verified under the conditions $(C1)$ and $(C2)$ that for any given bounded function $x(t),$ there exists a unique bounded on $\mathbb R$ solution $\phi_{x(t)}(t)$ of system (\ref{new_replicator}), which satisfies the following relation,
\begin{eqnarray}\label{integral_eqns}
\phi_{x(t)}(t)=\displaystyle\int^{t}_{-\infty} e^{A(t-s)}  [G(s,\phi_{x(t)}(s))  + H(x(s))] ds.  
\end{eqnarray}
Moreover, for a fixed $x(t),$ if $y(t)$ is a solution of (\ref{new_replicator}) with $y(0)=y_0 \in \mathbb R^n,$ then the inequality $\left\| \phi_{x(t)}(t)-y(t) \right\| \le N\left\| \phi_{x(t)}(0)-y_0 \right\| e^{(NL_1 - \omega) t}$ holds for $t\ge 0$ so that $\left\| \phi_{x(t)}(t)-y(t) \right\| \to 0$ as $t \to \infty.$  In other words, the bounded on $\mathbb R$ solution $\phi_{x(t)}(t)$ attracts all other solutions of (\ref{new_replicator}).

\section{Chaos Through a Cascade of Almost Periodic Solutions} \label{ap_cascade_section}

In this section, we will deal with the presence of chaos in the dynamics of (\ref{new_replicator}) through a cascade of almost periodic solutions. For that purpose, we will take advantage of a period-doubling cascade \cite{Feigenbaum80,Sander11,Sander12,Franceschini80,Sato83} of a system which will be used as the source for the perturbation function $x(t)$ in (\ref{new_replicator}). We will understand chaos in system (\ref{new_replicator}) as the presence of sensitivity, which is the main ingredient of chaos \cite{Dev90,Lorenz63,Wiggins88}, and the existence of infinitely many unstable almost periodic solutions in a bounded region.

Let us consider the system
\begin{eqnarray}
x'=F(t,x,\mu), \label{almost_periodicity_period_doubling}
\end{eqnarray}
where $\mu$ is a parameter and the function $F: \mathbb R \times \mathbb R^m \times \mathbb R \to \mathbb R^m$ is continuous in all of its arguments.

Suppose that there exists a positive number $T$ such that $F(t+T,x,\mu)=F(t,x,\mu)$ for all $t\in \mathbb R,$ $x\in \mathbb R^m$ and $\mu \in \mathbb R.$ We assume that system (\ref{almost_periodicity_period_doubling})  admits a period-doubling cascade \cite{Feigenbaum80,Sander11,Sander12}, that is, there exist a finite number $\mu_{\infty}$ and a sequence of period-doubling bifurcation values $\left\{\mu_j\right\},$ $j\in \mathbb N,$ satisfying $\mu_j \to \mu_{\infty}$ as $j \to \infty$ such that for each $j$ as the parameter $\mu$ increases or decreases through $\mu_j,$ system $(\ref{almost_periodicity_period_doubling})$ undergoes a period-doubling bifurcation, which gives rise to the formation of a new periodic solution with a twice period of the former one. At the parameter value $\mu=\mu_{\infty},$ there exist infinitely many unstable periodic solutions of (\ref{almost_periodicity_period_doubling}) all lying in a bounded region.  

The perturbation function $x(t)$ in (\ref{new_replicator}) will be provided by system (\ref{almost_periodicity_period_doubling}) with $\mu=\mu_{\infty},$ that is, it will be a solution of the system
\begin{eqnarray}\label{almost_periodicity_pd_generator}
x'=F(t,x,\mu_{\infty}). 
\end{eqnarray}

It is worth noting that the results of the present section are valid even if we replace the non-autonomous system (\ref{almost_periodicity_pd_generator}) with the autonomous equation 
\begin{eqnarray}\label{ap_autonomous_generator_pdc2}
x'=\overline{F}(x,\mu_{\infty}),
\end{eqnarray}
where $\overline{F}:\mathbb R^{m} \times \mathbb  R \to \mathbb R^{m}$  is a continuous function in all of its arguments.
 
We suppose that system (\ref{almost_periodicity_pd_generator}) ((\ref{ap_autonomous_generator_pdc2})) admits a chaotic attractor, let us say a set in $\mathbb R^m$ for (\ref{ap_autonomous_generator_pdc2}). Fix $x_0$ from the attractor, and take a solution $x(t)$ of (\ref{ap_autonomous_generator_pdc2}) with $x(0)=x_0.$ Since $x_0$ is a member of the chaotic attractor, we will call $x(t)$ a chaotic solution \cite{Dev90,Feigenbaum80,Lorenz63,Wiggins88}. 

There exists a compact set $\Lambda \subset \mathbb R^m$ such that the trajectories of the chaotic solutions of (\ref{almost_periodicity_pd_generator}) ((\ref{ap_autonomous_generator_pdc2})) lie inside $\Lambda$ for all $t.$ If we denote $M_H=\displaystyle \max_{x \in \Lambda} \left\|H(x)\right\|,$ then one can verify that $\displaystyle \sup_{t \in \mathbb R} \left\|\phi_{x(t)}(t)\right\| \leq \displaystyle \frac{N(M_G+M_H)}{\omega}$ for each chaotic solution $x(t)$ of (\ref{almost_periodicity_pd_generator}) ((\ref{ap_autonomous_generator_pdc2})).

The following conditions are needed throughout the section.
\begin{enumerate}
\item[\bf (C3)] There exists a positive number $L_2$ such that $\left\|H(x_1)-H(x_2)\right\| \le L_2 \left\|x_1-x_2\right\|$ for all $x_1,$ $x_2 \in \Lambda;$
\item[\bf (C4)] There exists a positive number $L_3$ such that $\left\|H(x_1)-H(x_2)\right\| \ge L_3 \left\|x_1-x_2\right\|$ for all $x_1,$ $x_2 \in \Lambda;$
\item[\bf (C5)] There exists a positive number $L_4$ such that $\left\|F(t,x_1,\mu_{\infty})-F(t,x_2,\mu_{\infty})\right\| \leq L_4\left\|x_1-x_2\right\|$ for all $t\in \mathbb R,$ $x_1,x_2\in \Lambda.$
\end{enumerate}

The next subsection is concerned with the extension of sensitivity.

\subsection{Extension of Sensitivity}

Let us describe the sensitivity feature for system (\ref{almost_periodicity_pd_generator}) as well as its replication by system (\ref{new_replicator}).

System (\ref{almost_periodicity_pd_generator}) is called sensitive if there exist positive numbers $\epsilon_0$ and $\Delta$ such that for an arbitrary positive number $\delta_0$ and for each chaotic solution $x(t)$ of (\ref{almost_periodicity_pd_generator}), there exist a chaotic solution $\overline{x}(t)$ of the same system, $t_0\in\mathbb R$ and an interval $J \subset [t_0,\infty)$ with a length no less than $\Delta$ such that $\left\|x(t_0)-\overline{x}(t_0)\right\|<\delta_0$ and $\left\|x(t)-\overline{x}(t)\right\| > \epsilon_0$ for all $t \in J.$ 

We say that system (\ref{new_replicator}) replicates the sensitivity of (\ref{almost_periodicity_pd_generator}) if there exist positive numbers $\epsilon_1$ and $\overline{\Delta}$ such that for an arbitrary positive number $\delta_1$ and  for each chaotic solution $x(t)$ of (\ref{almost_periodicity_pd_generator}), there exist a chaotic solution $\overline{x}(t)$ of (\ref{almost_periodicity_pd_generator}), $t_0\in\mathbb R$ and an interval $J^1\subset [t_0,\infty)$ with a length no less than $\overline{\Delta}$ such that $\left\|\phi_{x(t)}(t_0)-\phi_{\overline{x}(t)}(t_0)\right\|<\delta_1$ and $\left\|\phi_{x(t)}(t)-\phi_{\overline{x}(t)}(t)\right\| > \epsilon_1$ for all $t \in J^1.$ 
 
The next assertion is about the sensitivity feature of system (\ref{new_replicator}). 

\begin{lemma}\label{almost_periodicity_sensitivity_lemma}
Under the conditions $(C1)-(C5),$ system (\ref{new_replicator}) replicates the sensitivity of (\ref{almost_periodicity_pd_generator}). 
\end{lemma}

We omit the proof of Lemma \ref{almost_periodicity_sensitivity_lemma} since it can be proved in a very similar way to Lemma $5.1$ \cite{Akh8}.

We will handle the presence of chaos through a cascade of almost periodic motions in system (\ref{new_replicator}) in the following subsection.

\subsection{Cascade of Almost Periodic Solutions}

We say that system (\ref{new_replicator}) is chaotic through a cascade of almost periodic solutions if for each periodic solution $x(t)$ of (\ref{almost_periodicity_pd_generator}), system (\ref{new_replicator}) possesses an almost periodic solution.

One can verify by using the results of \cite{Fink74} that if $x(t)$ is a periodic solution of (\ref{almost_periodicity_pd_generator}), then the function $G(t,y)+H(x(t))$ in (\ref{new_replicator}) is almost periodic in $t$ uniformly for $y \in \mathbb R^n$ and the unique bounded on $\mathbb R$ solution $\phi_{x(t)}(t)$ of (\ref{new_replicator}) is almost periodic. On the other hand, if $x_1(t)$ and $x_2(t)$ are two different periodic solutions of (\ref{almost_periodicity_pd_generator}), then the corresponding almost periodic solutions $\phi_{x_1(t)}(t)$ and $\phi_{x_2(t)}(t)$ are different from each other under the condition $(C4).$ Because system (\ref{almost_periodicity_pd_generator}) is chaotic through period-doubling cascade, there exists a cascade of almost periodic solutions in the dynamics of (\ref{new_replicator}). Consequently, system (\ref{new_replicator}) possesses infinitely many almost periodic solutions in a bounded region. The instability of the existing almost periodic motions is ensured by Lemma \ref{almost_periodicity_sensitivity_lemma}. This result is mentioned in the following theorem.

\begin{theorem} \label{thm_pdc}
Under the conditions $(C1)-(C5),$ system (\ref{new_replicator}) is chaotic through a cascade of almost periodic solutions.
\end{theorem}

A corollary of Theorem \ref{thm_pdc} is as follows.

\begin{corollary}
Under the conditions $(C1)-(C5),$ the coupled system $(\ref{almost_periodicity_pd_generator})+(\ref{new_replicator})$ is chaotic through a cascade of almost periodic solutions.
\end{corollary}

In the next section, we will consider the formation of Li-Yorke chaos with infinitely many almost periodic motions.

\section{Li-Yorke Chaos with Infinitely Many Almost Periodic Motions} \label{ap_Li-Yorke_Section}

In the original paper  \cite{Li75}, chaos with infinitely many periodic solutions, which are separated from the elements of a scrambled set, was introduced. We modify the Li-Yorke definition of chaos by replacing periodic motions by almost periodic ones, and prove its presence in system (\ref{new_replicator}) rigorously.  

In opposition to the descriptions of regular motions such as periodic, quasi-periodic and almost periodic motions, one encounters with interaction of motions in order to describe the Li-Yorke chaos. Therefore, we need to introduce the concept of chaotic sets of functions.  

Let $\Lambda$ be a compact subset of $\mathbb R^m,$ and consider the set of uniformly bounded functions $\mathscr{A}$ whose elements are of the form $x(t):\mathbb R \to \Lambda.$ We suppose that $\mathscr{A}$ is an equicontinuous family on $\mathbb R.$ In this section, the perturbation function $x(t)$ in system (\ref{new_replicator}) will be provided from the elements of the collection $\mathscr{A}.$

A couple of functions $ \left( x(t), \overline{x}(t) \right) \in \mathscr{A} \times \mathscr{A}$ is called proximal if for an arbitrary small number $\epsilon>0$ and an arbitrary large number $E>0$ there exists an interval $J \subset \mathbb R$ with a length no less than $E$ such that $\left\|x(t)-\overline{x}(t)\right\| < \epsilon$ for all $t \in J.$ Besides, a couple of functions $\left( x(t), \overline{x}(t) \right) \in \mathscr{A} \times \mathscr{A}$ is frequently $(\epsilon_0, \Delta)-$separated if there exist numbers $\epsilon_0>0,$ $\Delta > 0$ and infinitely many disjoint intervals each with a length no less than $\Delta$ such that $\left\| x(t)-\overline{x}(t)\right\| > \epsilon_0$ for each $t$ from these intervals. 
It is worth noting that the numbers $\epsilon_0$ and $\Delta$ depend on the functions $x(t)$ and $\overline{x}(t).$
 
We say that a couple of functions $\left(x(t), \overline{x}(t) \right) \in \mathscr{A} \times \mathscr{A}$ is a Li-Yorke pair if they are proximal and frequently $(\epsilon_0, \Delta)-$separated for some positive numbers $\epsilon_0$ and $\Delta$.

The definition of a Li-Yorke chaotic set with infinitely many almost periodic motions is as follows.

\begin{definition} \label{almost_periodicity_defn_liyorke}
$\mathscr{A}$ is called a Li-Yorke chaotic set with infinitely many almost periodic motions if: 
\begin{itemize}
\item[i.] There exists a countably infinite set $\mathscr{R} \subset \mathscr{A}$ of almost periodic functions;
\item[ii.] There exists an uncountable set $\mathscr{D} \subset \mathscr{A},$ the scrambled set, such that the intersection of $\mathscr{D}$ and $\mathscr{R}$ is empty and each couple of different functions inside $\mathscr{D} \times \mathscr{D}$ is a Li-Yorke pair;
\item[iii.] For any function $x(t)\in \mathscr{D}$ and any almost periodic function $\overline{x}(t)\in \mathscr{R},$ the couple $\left(x(t),\overline{x}(t)\right)$ is frequently $(\epsilon_0, \Delta)-$separated for some positive numbers $\epsilon_0$ and $\Delta.$
\end{itemize}
\end{definition}

To provide a rigorous study of the subject, we introduce the following set of functions,
\begin{eqnarray}\label{A_y}
\mathscr{B}=\left\{\phi_{x(t)}(t) ~|~  x(t) \in \mathscr{A} \right\}.
\end{eqnarray}


The following assertions, which are about the proximality and frequent separation features for system (\ref{new_replicator}), can be proved in a similar way to Lemma $6.1$ and Lemma $6.2$ \cite{Akh8}.

\begin{lemma} \label{almost_periodicity_proximality}
Under the conditions $(C1)-(C3),$ if a couple of functions $\left(x(t),\overline{x}(t)\right) \in \mathscr{A} \times \mathscr{A} $ is proximal, then the same is true for the couple $\left(\phi_{x(t)}(t),\phi_{\overline{x}(t)}(t)\right) \in \mathscr{B} \times \mathscr{B}.$
\end{lemma} 
 
\begin{lemma} \label{almost_periodicity_frequent_separation}
Assume that the conditions $(C1),$ $(C2)$ and $(C4)$ hold. If a couple of functions $\left(x(t),\overline{x}(t)\right) \in \mathscr{A} \times \mathscr{A} $ is frequently $(\epsilon_0,\Delta)-$separated for some positive numbers $\epsilon_0$ and $\Delta$, then the couple of functions $\left(\phi_{x(t)}(t),\phi_{\overline{x}(t)}(t) \right) \in \mathscr{B} \times  \mathscr{B} $ is frequently $(\epsilon_1,\overline{\Delta})-$separated for some positive numbers $\epsilon_1$ and $\overline{\Delta}.$
\end{lemma}

The main result of the present section is mentioned in the following theorem.

\begin{theorem} \label{ap_main_thm_li_yorke}
Suppose that the conditions $(C1)-(C4)$ are fulfilled. If the set $\mathscr{A}$ is chaotic in the sense of Definition \ref{almost_periodicity_defn_liyorke}, then the same is true for the set $\mathscr{B}.$
\end{theorem}

\noindent \textbf{Proof.} 
Let $\mathscr{R}$ be the set of almost periodic functions inside $\mathscr{A}.$ If $x(t)$ belongs to $\mathscr{R},$ then the function $G(t,y)+H(x(t))$ in (\ref{new_replicator}) is almost periodic in $t$ uniformly for $y \in \mathbb R^n$ \cite{Fink74}. In this case, the bounded solution $\phi_{x(t)}(t)$ of (\ref{new_replicator}) is almost periodic \cite{Fink74}. Consider the set $\widetilde{\mathscr{R}} = \left\{ \phi_{x(t)}(t) ~|~ x(t) \in \mathscr{R} \right\}.$ According to condition $(C4)$ there is a one-to-one correspondence between $\mathscr{R}$ and $\widetilde{\mathscr{R}}.$ Since the set $\mathscr{R}$ is countably infinite, the set $\widetilde{\mathscr{R}}$ is also countably infinite.  

Suppose that $\mathscr{D}$ is a scrambled set inside $\mathscr{A}.$ Define the set 
$$\widetilde{\mathscr{D}} = \left\{ \phi_{x(t)}(t) ~|~ x(t) \in \mathscr{D} \right\}.$$ The set $\widetilde{\mathscr{D}}$ is uncountable since the same is true for $\mathscr{D}.$ Moreover, the intersection of $\widetilde{\mathscr{D}}$ and $\widetilde{\mathscr{R}}$ is empty since the intersection of $\mathscr{D}$ and $\mathscr{R}$ is empty. 

Because each couple of different functions inside $\mathscr{D} \times \mathscr{D}$ is proximal, Lemma $\ref{almost_periodicity_proximality}$ implies the same feature for each couple of different functions inside $\widetilde{\mathscr{D}} \times \widetilde{\mathscr{D}}.$ On the other hand, since each couple of different functions $\left(x(t),\overline{x}(t)\right) \in \mathscr{D} \times \mathscr{D}$ is frequently $(\epsilon_0,\Delta)-$separated for some positive numbers $\epsilon_0$ and $\Delta$, Lemma $\ref{almost_periodicity_frequent_separation}$ implies that each couple of different functions $\left(y(t),\overline{y}(t) \right) \in \widetilde{\mathscr{D}} \times \widetilde{\mathscr{D}}$ is frequently $(\epsilon_1,\overline{\Delta})-$separated for some positive numbers $\epsilon_1$ and $\overline{\Delta}.$ Additionally, the frequent separation feature holds also for each pair inside $\widetilde{\mathscr{D}} \times \widetilde{\mathscr{R}}.$ Therefore, the set $\widetilde{\mathscr{D}}$ is a scrambled set inside $\mathscr{B}.$ Consequently, $\mathscr{B}$ is a Li-Yorke chaotic set in accordance with Definition \ref{almost_periodicity_defn_liyorke}. $\square$

Using the technique of the proof of Theorem \ref{ap_main_thm_li_yorke}, one can confirm that if the function $G(t,y),$ which is used in the right hand side of system $(\ref{new_replicator}),$ is quasi-periodic (periodic) in $t$ uniformly for $y \in \mathbb R^n$, then under the conditions of Theorem \ref{ap_main_thm_li_yorke}, the set $\mathscr{B}$ is chaotic in the sense of Li-Yorke with infinitely many quasi-periodic (periodic) motions provided that the same is true for $\mathscr{A}.$

In the following section, the theoretical results of Section \ref{ap_Li-Yorke_Section} will be supported by simulations. We will show how Theorem \ref{ap_main_thm_li_yorke} can be utilized to obtain a Li-Yorke chaotic Duffing oscillator with infinitely many almost periodic motions. In order to construct the Li-Yorke chaotic set $\mathscr{A},$ we will take into account the bounded solutions of a Duffing oscillator perturbed with a relay function whose switching moments are changing chaotically.

\section{An Example} \label{almost_periodicity_example}

Let us consider the forced Duffing oscillator 
\begin{eqnarray} \label{ap_ex1}
x''+x'+\frac{11}{2} x+0.0003 x^3=\sin(2\pi t) + \nu(t,\theta_0),
\end{eqnarray}
where $\nu(t,\theta_0)$ is the relay function defined as
\begin{eqnarray} \label{ap_nu}
\mathcal \nu(t,\theta_0)=\left\{\begin{array}{ll} 3.6, ~\textrm{if}  & \theta_{2i} < t  \leq \theta_{2i+1}, \\
                                                  1.3, ~\textrm{if}  & \theta_{2i-1} < t \leq \theta_{2i}. 
\end{array} \right.
\end{eqnarray} 
In the relay function (\ref{ap_nu}), the sequence $\left\{\theta_i\right\},$ $i\in\mathbb Z,$ of switching moments is defined through the equation $\theta_i=i+\kappa_i,$ in which the sequence $\left\{\kappa_i\right\},$ $\kappa_0 \in [0,1],$ is generated by the logistic map such that  
\begin{eqnarray} \label{ap_logistic_map}
\kappa_{i+1}=\lambda \kappa_i (1-\kappa_i),
\end{eqnarray}
where $\lambda$ is a parameter. The interval $[0,1]$ is invariant under the iterations of (\ref{ap_logistic_map}) for the values of $\lambda$ between $0$ and $4$ \cite{Hale91}. For $\lambda=3.9,$ the map (\ref{ap_logistic_map}) is Li-Yorke chaotic \cite{Li75}, and the family $\left\{\nu(t,\theta_0)\right\},$ $\theta_0\in[0,1],$ is chaotic in the sense of Li-Yorke with infinitely many periodic motions \cite{Akh3}.

Using the variables $x_1=x$ and $x_2=x',$ equation (\ref{ap_ex1}) can be written as a system in the form
\begin{eqnarray}
\begin{array}{l} \label{ap_ex2}
x_1'=x_2 \\
x_2'=- \displaystyle \frac{11}{2} x_1 -x_2 -0.0003 x_1^3 + \sin(2\pi t) + \nu(t,\theta_0).
\end{array}
\end{eqnarray}

According to the results of the paper \cite{Akh3}, the set $\mathscr{A}$ consisting of the bounded on $\mathbb R$ solutions of (\ref{ap_ex2}) with $\lambda=3.9$ is Li-Yorke chaotic such that it possesses infinitely many periodic solutions with periods $2p,$ $p\in\mathbb N,$ which are separated from the motions of the scrambled set. The chaoticity of the switching moments $\left\{\theta_i\right\}$ gives rise to the presence of chaos in system (\ref{ap_ex2}). The reader is referred to \cite{Akh4,Akh3,Akh5,Akh7,Akh17,Akh21,Akh22} for further information about the dynamics of relay systems. 

\begin{figure}[ht] 
\centering
\includegraphics[width=12.0cm]{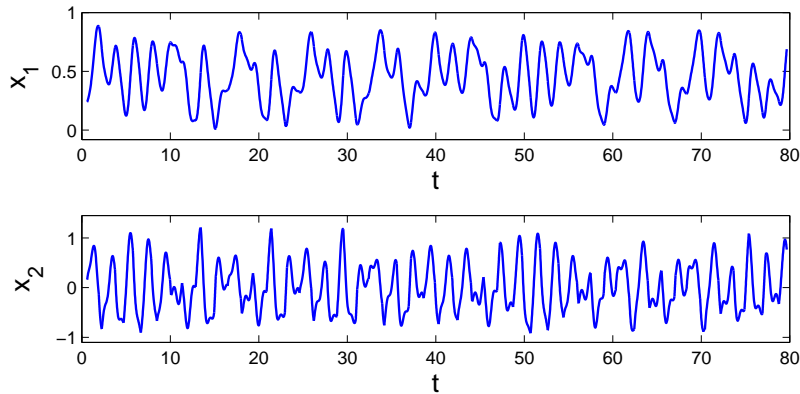}
\caption{Li-Yorke chaos in system (\ref{ap_ex2}).}
\label{ap_fig1}
\end{figure}

In what follows, we will take into account system (\ref{ap_ex2}) with $\lambda=3.9.$ To illustrate the chaotic dynamics of (\ref{ap_ex2}), let us consider the solution of the system with the initial data $x_1(t_0)=0.24,$ $x_2(t_0)=0.16,$ where $t_0=0.61.$ The simulation results are shown in Figure \ref{ap_fig1}, where one can observe the chaotic behavior of (\ref{ap_ex2}). The sequence $\left\{\theta_i\right\}$ of switching moments is used with $\theta_0=0.61$ in the simulation. It is worth noting that the bounded on $\mathbb R$ solutions of (\ref{ap_ex2}) lie inside the compact region $\Lambda=\left\{(x_1,x_2) \in \mathbb R^2: 0 \le x_1 \le 0.9, \ -1 \le x_2 \le 1.3 \right\}.$

Next, let us consider the quasi-periodically forced Duffing oscillator
\begin{eqnarray} \label{quasiperiodically_forced_duffing}
u''+\frac{3}{2} u' + 2u + 0.0001u^3 = \frac{3}{2}\sin\left( \frac{t}{\sqrt{2}} \right)+\frac{1}{2} \sin(\pi t).
\end{eqnarray}
By means of the variables $u_1=u$ and $u_2=u'$ one can reduce (\ref{quasiperiodically_forced_duffing}) to the system 
\begin{eqnarray} \label{ap_ex4}
\begin{array}{l}
u_1'=u_2 \\
\displaystyle u_2'=-2u_1-\frac{3}{2}u_2-0.0001u_1^3+\frac{3}{2}\sin\left( \frac{t}{\sqrt{2}} \right)+\frac{1}{2} \sin(\pi t).
\end{array} 
\end{eqnarray}
We perturb (\ref{ap_ex4}) with the solutions of (\ref{ap_ex2}), and set up the following system,
\begin{eqnarray} \label{ap_ex3}
\begin{array}{l}
y_1'=y_2 + 6 \arctan(x_1(t)) \\
\displaystyle y_2'=-2y_1-\frac{3}{2}y_2-0.0001 y_1^3 +\frac{3}{2}\sin\left( \frac{t}{\sqrt{2}} \right)+\frac{1}{2} \sin(\pi t) + 2x_2(t).
\end{array} 
\end{eqnarray}

System (\ref{ap_ex3}) is in the form of (\ref{new_replicator}) with 
$$A=\left(
\begin {array}{ccc}
0&1\\
\noalign{\medskip}
-2&-\displaystyle\frac{3}{2}
\end {array}
\right),$$ 
$$G(t,y_1,y_2)=\left(
\begin {array}{ccc}
0\\
\noalign{\medskip}
\displaystyle -0.0001 y_1^3 +\frac{3}{2}\sin\left( \frac{t}{\sqrt{2}} \right)+\frac{1}{2} \sin(\pi t)
\end {array}
\right),$$ and
$$H(x_1,x_2)=\left(
\begin {array}{ccc}
6\arctan(x_1)\\
\noalign{\medskip}
\displaystyle  2x_2
\end {array}
\right).$$ 
The eigenvalues of the matrix $A$ are $- \displaystyle \frac{3}{4} \pm i \frac{\sqrt{23}}{4},$ and the function $G(t,y_1,y_2)$ is quasi-periodic in $t$ uniformly for $(y_1,y_2)\in\mathbb R^2.$ One can confirm that 
$$
e^{At} = e^{-3t/4} P \left(
\begin {array}{ccc}
\cos\bigg(\displaystyle \frac{\sqrt{23}}{4}t\bigg)&-\sin\bigg(\displaystyle \frac{\sqrt{23}}{4}t\bigg)\\
\noalign{\medskip}
\sin\bigg(\displaystyle \frac{\sqrt{23}}{4}t\bigg)&\cos\bigg(\displaystyle \frac{\sqrt{23}}{4}t\bigg)
\end {array}
\right)P^{-1},
$$
where
$$P=\left(
\begin {array}{ccc}
0&1\\
\noalign{\medskip}
\displaystyle\frac{\sqrt{23}}{4} &-\displaystyle\frac{3}{4}
\end {array}
\right).$$ 
Therefore, $\left\|e^{At}\right\| \le N e^{-\omega t},$ $t \ge 0,$ where $N = \left\|P\right\| \left\|P^{-1}\right\| \approx 2.0029$ and $\omega = 3/4.$

\begin{figure}[ht] 
\centering
\includegraphics[width=12.0cm]{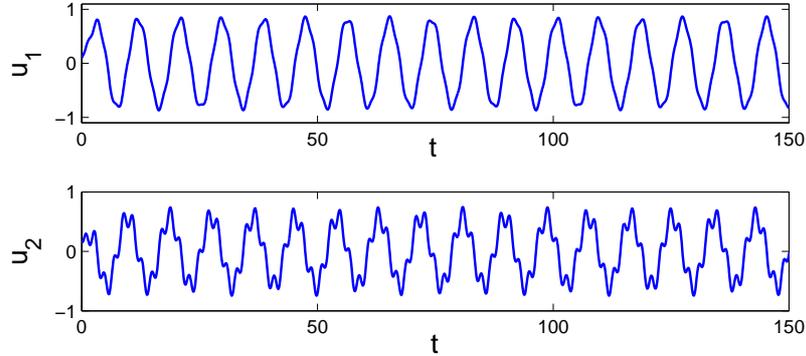}
\caption{The quasi-periodic behavior of system (\ref{ap_ex4}).}
\label{ap_fig2}
\end{figure}

In order to show that system (\ref{ap_ex4}) possesses an asymptotically stable quasi-periodic solution, let us consider the solution of (\ref{ap_ex4}) with initial data $u_1(0)=0.1,$ $u_2(0)=0.2.$ The $u_1$ and $u_2$ coordinates of the solution are depicted in Figure \ref{ap_fig2}, where one can observe that the represented solution approaches to the asymptotically stable quasi-periodic solution such that the system does not possess chaos.

It can be numerically verified that the bounded on $\mathbb R$ solutions of (\ref{ap_ex3}) lie inside the compact region $\mathscr{R}=\left\{(y_1,y_2) \in \mathbb R^2: -1.1 \le y_1 \le 4.7,  \ -5.3 \le y_2 \le 0.5 \right\}.$ Therefore, it is reasonable to consider the dynamics of (\ref{ap_ex3}) inside $\mathscr{R},$ and conditions $(C1)-(C4)$ are valid for system (\ref{ap_ex3}) with $M_G=2.0104,$ $L_1=0.006627,$ $L_2=6\sqrt{2},$ and $L_3=\sqrt{2}.$ According to the theoretical results of Section \ref{ap_Li-Yorke_Section}, the set $\mathscr{B}$ consisting of the bounded on $\mathbb R$ solutions of (\ref{ap_ex3}) for which $(x_1(t),x_2(t))$ belongs to $\mathscr{A}$ is Li-Yorke chaotic with infinitely many quasi-periodic solutions, which are separated from the motions of the scrambled set. That is, the applied perturbation $H(x_1(t),x_2(t))$ effects (\ref{ap_ex4}) in such a way that Li-Yorke chaos takes place in the dynamics of (\ref{ap_ex3}). 

\begin{figure}[ht] 
\centering
\includegraphics[width=12.0cm]{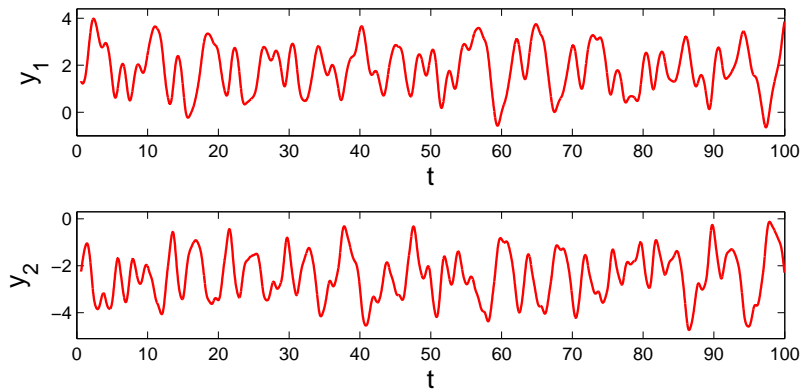}
\caption{The graphs of the $y_1$ and $y_2$ coordinates of system (\ref{ap_ex3}).}
\label{ap_fig3}
\end{figure}

In system (\ref{ap_ex3}) we use the solution of (\ref{ap_ex2}) that is depicted in Figure \ref{ap_fig1}, and we represent in Figure \ref{ap_fig3} the solution of (\ref{ap_ex3}) corresponding to the initial data $y_1(t_0)=1.32,$ $y_2(t_0)=-2.25,$ where $t_0=0.61.$ Figure \ref{ap_fig3} supports the result of Theorem \ref{ap_main_thm_li_yorke} such that system (\ref{ap_ex3}) possesses chaotic motions.

Now, we will demonstrate that the chaos formation mechanism can be proceeded further. For that purpose, we take into account the Duffing equation
\begin{eqnarray} \label{ap_third_Duffing}
v''+2v'+4v+0.0002v^3=0,
\end{eqnarray}
which is equivalent to the system
\begin{eqnarray}
\begin{array}{l} \label{ap_ex6}
v_1'=v_2 \\
v_2'=-4 v_1-2 v_2-0.0002 v_1^3, 
\end{array}
\end{eqnarray}
in view of the variables $v_1=v$ and $v_2=v'.$ One can verify that  (\ref{ap_ex6}) admits an asymptotically stable equilibrium point.

We perturb system (\ref{ap_ex6}) with the solutions of (\ref{ap_ex3}) to constitute the system
\begin{eqnarray}
\begin{array}{l} \label{ap_ex5}
\displaystyle z_1'=z_2+y_1(t)+\frac{1}{2}\sin(y_1(t)) \\
\displaystyle z_2'=-4 z_1-2 z_2-0.0002 z_1^3+\frac{1}{5} y_2(t)+ \frac{1}{10} y_2^3(t). 
\end{array}
\end{eqnarray}
It is worth noting that (\ref{ap_ex5}) is in the form of (\ref{new_replicator}) with 
$A=\left(
\begin {array}{ccc}
0&1\\
\noalign{\medskip}
-4&-2
\end {array}
\right),$
whose eigenvalues are $-1\pm i\sqrt{3}.$ 

It can be shown in a similar way to system (\ref{ap_ex3}) that conditions $(C1)-(C4)$ are valid also for (\ref{ap_ex5}). According to our results, the collection $\mathscr{C}$ of bounded on $\mathbb R$ solutions of (\ref{ap_ex5}) for which $(y_1(t),y_2(t))$ belongs to $\mathscr{B}$ is a Li-Yorke chaotic set with infinitely many quasi-periodic motions.

\begin{figure}[ht] 
\centering
\includegraphics[width=12.0cm]{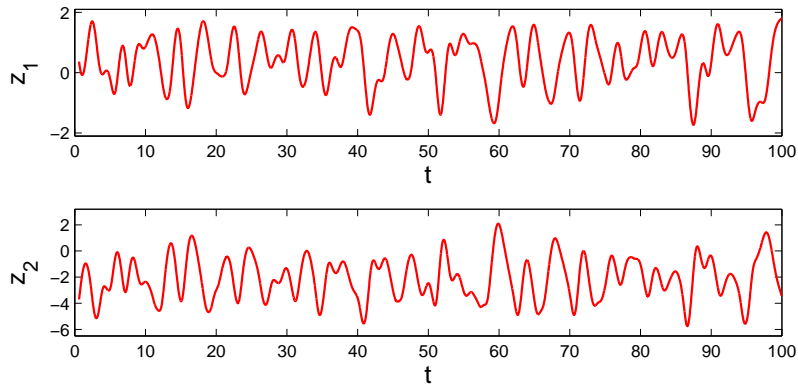}
\caption{The chaotic behavior of system (\ref{ap_ex5}).}
\label{ap_fig4}
\end{figure}

Utilizing the solution of (\ref{ap_ex3}) that is shown in Figure \ref{ap_fig3} as the perturbation in system (\ref{ap_ex5}), we depict in Figure \ref{ap_fig4} the solution of (\ref{ap_ex5}) with $z_1(t_0)=0.37,$ $z_2(t_0)=-3.71,$ where $t_0=0.61.$ Figure \ref{ap_fig4} reveals that the system (\ref{ap_ex5}) is Li-Yorke chaotic.
 
\begin{figure}[ht] 
\centering
\includegraphics[width=7.5cm]{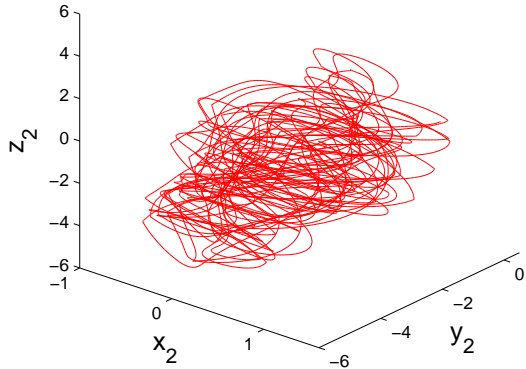}
\caption{The projection of the chaotic trajectory of system $(\ref{ap_ex2})+(\ref{ap_ex3})+(\ref{ap_ex5})$ on the $x_2-y_2-z_2$ space.}
\label{ap_fig5}
\end{figure}

To confirm one more time that the $6-$dimensional system  $(\ref{ap_ex2})+(\ref{ap_ex3})+(\ref{ap_ex5})$ admits a chaotic attractor, we represent in Figure \ref{ap_fig5} the projection of the trajectory of $(\ref{ap_ex2})+(\ref{ap_ex3})+(\ref{ap_ex5})$ corresponding to the initial data $x_1(t_0)=0.24,$ $x_2(t_0)=0.16,$ $y_1(t_0)=1.32,$ $y_2(t_0)=-2.25,$ $z_1(t_0)=0.37,$ $z_2(t_0)=-3.71,$ where $t_0=0.61,$ on the $x_2-y_2-z_2$ space. The simulation result confirms that system  $(\ref{ap_ex2})+(\ref{ap_ex3})+(\ref{ap_ex5})$ possesses a chaotic attractor.

\section{Control of Tori} \label{ap_control_section}

In this section, we will numerically demonstrate the stabilization of the unstable quasi-periodic motions embedded in the chaotic attractor of the unidirectionally coupled Duffing oscillators $(\ref{ap_ex2})+(\ref{ap_ex3})+(\ref{ap_ex5}).$ We suppose that to stabilize the quasi-periodic solutions of (\ref{ap_ex3}) and (\ref{ap_ex5})  it is sufficient to control the chaos of system (\ref{ap_ex2}). For that purpose we will make use of the Ott-Grebogi-Yorke (OGY) control method \cite{Ott90} applied to the logistic map (\ref{ap_logistic_map}), which gives rise to the presence of chaos in (\ref{ap_ex2}). We proceed by briefly explaining the OGY control method for the map \cite{Sch99}.

Suppose that the parameter $\lambda$ in the logistic map (\ref{ap_logistic_map}) is allowed to vary in the range $[3.9-\varepsilon, 3.9+\varepsilon]$, where $\varepsilon$ is a given small number. Consider an arbitrary solution $\left\{\kappa_i\right\},$ $\kappa_0\in[0,1],$ of the map and denote by $\kappa^{(j)},$ $j=1,2,\ldots ,p,$ the target unstable $p-$periodic orbit to be stabilized. 
In the OGY control method \cite{Sch99}, at each iteration step $i$ after the control mechanism is switched on, we consider the map (\ref{ap_logistic_map}) with the parameter value $\lambda=\bar \lambda_i,$ where
\begin{eqnarray}\label{ap_formula}
\bar \lambda_i=3.9 \left(1+\frac{[2\kappa^{(j)}-1][\kappa_{i}-\kappa^{(j)}]}{\kappa^{(j)}[1-\kappa^{(j)}]} \right),
\end{eqnarray}
provided that the number on the right-hand side of the formula $(\ref{ap_formula})$ belongs to the interval $[3.9-\varepsilon, 3.9+\varepsilon].$ In other words, we apply a perturbation in the amount of $ \displaystyle\frac{3.9[2\kappa^{(j)}-1][\kappa_{i}-\kappa^{(j)}]}{\kappa^{(j)}[1-\kappa^{(j)}]}$ to the parameter $\lambda=3.9$ of the logistic map, if the trajectory $\left\{\kappa_i\right\}$ is sufficiently close to the target periodic orbit. This perturbation makes the map behave regularly so that at each iteration step the orbit $\kappa_i$ is forced to be located in a small neighborhood of a previously chosen periodic orbit $\kappa^{(j)}.$ Unless the parameter perturbation is applied, the orbit $\kappa_i$ moves away from $\kappa^{(j)}$ due to the instability. If $\displaystyle \left|\frac{ 3.9[2\kappa^{(j)}-1][\kappa_{i}-\kappa^{(j)}]}{\kappa^{(j)}[1-\kappa^{(j)}]} \right| > \varepsilon$, we set $\bar \lambda_{i}=3.9,$ so that the system evolves at its original parameter value, and wait until the trajectory $\left\{\kappa_i\right\}$ enters a sufficiently small neighborhood of the periodic orbit $\kappa^{(j)},$ $j=1,2,\ldots ,p,$ such that the inequality $-\varepsilon \le \displaystyle\frac{3.9 [2\kappa^{(j)}-1][\kappa_{i}-\kappa^{(j)}]}{\kappa^{(j)}[1-\kappa^{(j)}]}  \le \varepsilon$ holds. If this is the case, the control of chaos is not achieved immediately after switching on the control mechanism. Instead, there is a transition time before the desired periodic orbit is stabilized. The transition time increases if the number $\varepsilon$ decreases \cite{Gon04}.

\begin{figure}[ht] 
\centering
\includegraphics[width=12.7cm]{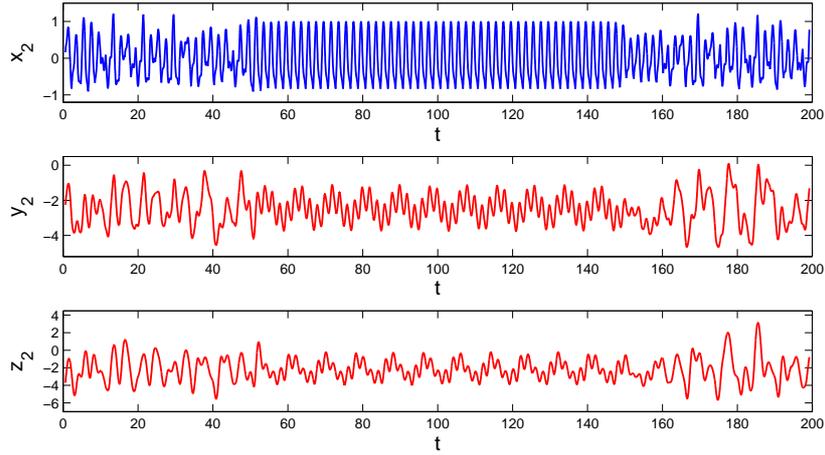}
\caption{The application of the OGY control method around the fixed point $2.9/3.9$ of the logistic map (\ref{ap_logistic_map}) for the stabilization of quasi-periodic motions of the coupled Duffing oscillators $(\ref{ap_ex2})+(\ref{ap_ex3})+(\ref{ap_ex5}).$ The value $\varepsilon=0.06$ is used in the simulation. Control is switched on at $t=\theta_{35}$ and switched off at $t=\theta_{95}.$}
\label{ap_fig6}
\end{figure}

\begin{figure}[ht] 
\centering
\includegraphics[width=7.5cm]{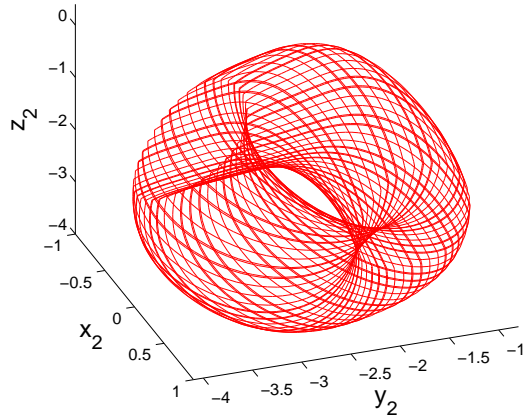}
\caption{The stabilized torus of system $(\ref{ap_ex2})+(\ref{ap_ex3})+(\ref{ap_ex5}).$}
\label{ap_fig7}
\end{figure}

An unstable $2-$periodic solution of (\ref{ap_ex2}) can be stabilized by controlling the $1-$periodic orbit of the logistic map (\ref{ap_logistic_map}), i.e.  the fixed point $2.9/3.9$ of the map. We apply the OGY control method around the fixed point of the logistic map to stabilize the corresponding quasi-periodic solution of the $6-$dimensional system $(\ref{ap_ex2})+(\ref{ap_ex3})+(\ref{ap_ex5}).$ Figure \ref{ap_fig6} represents the $x_2,$ $y_2$ and $z_2$ coordinates of $(\ref{ap_ex2})+(\ref{ap_ex3})+(\ref{ap_ex5})$ corresponding to the initial data  $x_1(t_0)=0.24,$ $x_2(t_0)=0.16,$ $y_1(t_0)=1.32,$ $y_2(t_0)=-2.25,$ $z_1(t_0)=0.37,$ $z_2(t_0)=-3.71,$ where $t_0=0.61.$ The control mechanism is switched on at $t=\theta_{35}$ and switched off at $t=\theta_{95}.$ The value $\varepsilon=0.06$ is used in the simulation. There is a transition time such that the control becomes dominant approximately at $t=53,$ and it prolongs approximately till $t=148,$ after which chaos develops again. It is seen in Figure \ref{ap_fig6} that the $2-$periodic solution of system (\ref{ap_ex2}) is controlled for $53 \le t \le 148,$ and accordingly the corresponding quasi-periodic solutions of systems (\ref{ap_ex3}) and (\ref{ap_ex5}) are stabilized in the same interval of time. Moreover, we illustrate the stabilized torus of system $(\ref{ap_ex2})+(\ref{ap_ex3})+(\ref{ap_ex5})$ in Figure \ref{ap_fig7}. Both of the Figures \ref{ap_fig6} and \ref{ap_fig7} support the result of Theorem \ref{ap_main_thm_li_yorke} such that they manifest the presence of quasi-periodic motions embedded in the chaotic attractor of $(\ref{ap_ex2})+(\ref{ap_ex3})+(\ref{ap_ex5}).$ 

\section{Conclusions} \label{ap_conclusion}

The possibility of replacing infinitely many periodic solutions by more general types of regular motions in a chaotic attractor was considered in the studies \cite{Shilnikov02,Seifert97,Nemytskii60}. The present paper provides a method to obtain chaotic attractors with infinitely many almost periodic motions instead of periodic ones. In this way, the complexity of chaos increases. 

Chaos in the sense of Li-Yorke and the one obtained through a cascade of almost periodic motions are considered in the present study. One of the advantages of the proposed procedure is the controllability of the obtained chaos. A control technique for stabilizing the unstable almost periodic motions is presented, and control of tori is numerically demonstrated by means of the OGY control method \cite{Ott90} in unidirectionally coupled Duffing oscillators. The results reveal that the OGY control method is suitable to stabilize not only periodic motions but also almost periodic ones. Other chaos control methods such as the Pyragas method \cite{Pyragas92} can also be used for that purpose. It is worth noting that the presented method is appropriate to obtain route to chaos by intermittency \cite{Pomeau80} in which regular behavior of almost periodic type is interrupted by sporadic bursts of chaotic behavior.

According to Watanabe et al. \cite{Watanabe97} chaos increases the capacity of memorizing in neural networks. One can suppose that chaos with almost periodic motions provides a memory with a larger capacity than that with periodic motions. Therefore, the obtained results may be useful for the theory of neural networks and investigations of brain activities.


\begin{thebibliography}{999}

\bibitem{Dev90} 
Devaney, R. (1987), \textit{An Introduction to Chaotic Dynamical Systems}, Addison-Wesley: United States of America.

\bibitem{Li75} 
Li, T.Y. and Yorke, J.A. (1975), Period three implies chaos, \textit{The American Mathematical Monthly}, \textbf{82}, 985-992.

\bibitem{Feigenbaum80} 
Feigenbaum, M.J. (1980), Universal behavior in nonlinear systems, \textit{Los Alamos Science}, \textbf{1}, 4-27.

\bibitem{Gonchenko96} 
Gonchenko, S.V., Shil'nikov L.P. and Turaev, D.V. (1996), Dynamical phenomena in systems with structurally unstable Poincar\'{e} homoclinic orbits, \textit{Chaos}, \textbf{6}, 15-31.

\bibitem{Shilnikov67} 
Shil'nikov, L.P. (1967), On a Poincar\'{e}-Birkhoff problem, \textit{Math. USSR-Sbornik}, \textbf{3}, 353-371.

\bibitem{Smale65} 
Smale, S. (1965), \textit{Diffeomorphisms with many periodic points}, Differential and Combinatorial Topology: A Symposium in Honor of Marston Morse, Princeton University Press: Princeton, NJ, pp. 63-70.

\bibitem{Shilnikov02} 
Shilnikov, L. (2002), \textit{Bifurcations and strange attractors}, Proceedings of the International Congress of Mathematicians, vol. III., Higher Ed. Press: Beijing, pp. 349-372.

\bibitem{Seifert97} 
Seifert, G. (1997), On chaos in general semiflows, \textit{Nonlinear Analysis, Theory, Methods and Applications}, \textbf{28}, 1719-1727.

\bibitem{Ott90} 
Ott, E., Grebogi, C. and Yorke, J.A. (1990), Controlling chaos, \textit{Phys. Rev. Lett.}, \textbf{64}, 1196-1199.

\bibitem{Andersson94} 
Andersson, K.G. (1994), Poincar\'{e}'s discovery of homoclinic points, \textit{Archive for History of Exact Sciences}, \textbf{48}, 133-147.

\bibitem{Cartwright1} 
Cartwright, M. and Littlewood, J. (1945), On nonlinear differential equations of the second order I: The equation $\ddot{y}- k(1 - y^2)'y + y = bk cos(\lambda t + a),$ $k$ large, \textit{J. London Math. Soc.}, \textbf{20}, 180-189.

\bibitem{Levinson} 
Levinson, N. (1949), \textit{A second order differential equation with singular solutions}, Ann. of Math., \textbf{50}, 127-153.

\bibitem{Lorenz63} 
Lorenz, E.N. (1963), Deterministic nonperiodic flow, \textit{J. Atmos. Sci.}, \textbf{20}, 130-141.

\bibitem{Ueda78} 
Ueda, Y. (1978), Random phenomena resulting from non-linearity in the system described by Duffing's equation, \textit{Trans. Inst. Electr. Eng. Jpn.}, \textbf{98A}, 167-173.

\bibitem{Grebogi97} 
Grebogi, C. and  Yorke, J.A. (1997), \textit{The Impact of Chaos on Science and Society}, United Nations University Press: Tokyo.

\bibitem{Huang02} 
Huang, W. and Ye, X. (2002), Devaney's chaos or 2-scattering implies Li-Yorke's chaos, \textit{Topology Appl.}, \textbf{117}, 259-272.

\bibitem{Sharkovskii} 
Sharkovskii, A.N. (1964), Coexistence of cycles of a continuous map of the line into itself, \textit{Ukrainian Mathematical Journal}, \textbf{16}, 61-71 (In Russian). 

\bibitem{Myrberg58} 
Myrberg, P.J. (1958), Iteration von Quadratwurzeloperationen. I, \textit{Ann. Acad. Sci. Fenn. Ser. A}, \textbf{256}, 1-10.

\bibitem{Myrberg59} 
Myrberg, P.J. (1959), Iteration von Quadratwurzeloperationen. II, \textit{Ann. Acad. Sci. Fenn. Ser. A}, \textbf{268}, 1-10.

\bibitem{Myrberg63} 
Myrberg, P.J. (1963), Iteration von Quadratwurzeloperationen. III, \textit{Ann. Acad. Sci. Fenn. Ser. A}, \textbf{336}, 1-10.

\bibitem{Sander11} 
Sander, E. and Yorke, J.A. (2011), Period-doubling cascades galore, \textit{Ergod. Th. \& Dynam. Sys.}, \textbf{31}, 1249-1267.

\bibitem{Sander12} 
Sander, E. and Yorke, J.A. (2012), Connecting period-doubling cascades to chaos, \textit{Int. J. Bifurcation and Chaos}, \textbf{22}, 1250022. 

\bibitem{Barrio14} 
Barrio, R., Martinez, M.A., Serrano, S. and Shilnikov, A. (2014), Macro- and micro-chaotic structures in the Hindmarsh-Rose model of bursting neurons, \textit{Chaos}, \textbf{24}, 023128.

\bibitem{Lacitignola10} 
Lacitignola, D., Petrosillo, I. and  Zurlini, G. (2010), Time-dependent regimes of a tourism-based social-ecological system: Period-doubling route to chaos, \textit{Ecological Complexity}, \textbf{7}, 44-54.

\bibitem{Hanias09} 
Hanias, M.P., Avgerinos, Z. and Tombras, G.S. (2009), Period doubling, Feigenbaum constant and time series prediction in an experimental chaotic RLD circuit, \textit{Chaos, Solitons \& Fractals}, \textbf{40}, 1050-1059.

\bibitem{Simpson94} 
Simpson, T.B., Liu, J.M., Gavrielides, A., Kovanis, V. and Alsing, P.M. (1994), Period-doubling route to chaos in a semiconductor laser subject to optical injection, \textit{Applied Physics Letters}, \textbf{64}, 3539-3541.

\bibitem{Akh2} 
Akhmet, M.U. (2009), Li-Yorke chaos in the impact system, \textit{J. Math. Anal. Appl.}, \textbf{351}, 804-810. 

\bibitem{Akh4} 
Akhmet, M.U. (2009), Dynamical synthesis of quasi-minimal sets, \textit{Int. J. Bifurcation and Chaos}, \textbf{19}, 2423-2427. 

\bibitem{Akh3} 
Akhmet, M.U. (2009), Creating a chaos in a system with relay, \textit{International Journal of Qualitative Theory of Differential Equations and Applications}, \textbf{3}, 3-7.

\bibitem{Akh5} 
Akhmet, M.U. (2009), Devaney's chaos of a relay system, \textit{Commun. Nonlinear Sci. Numer. Simulat.}, \textbf{14}, 1486-1493.
 
\bibitem{Akh7} 
Akhmet, M.U. and Fen, M.O. (2012), Chaotic period-Doubling and OGY control for the forced Duffing equation, \textit{Commun. Nonlinear Sci. Numer. Simulat.}, \textbf{17}, 1929-1946. 
 
\bibitem{Akh8} 
Akhmet, M.U. and Fen, M.O. (2013), Replication of chaos, \textit{Commun. Nonlinear Sci. Numer. Simulat.}, \textbf{18}, 2626-2666. 
 
\bibitem{Akh17} 
Akhmet, M.U. and Fen, M.O. (2013), Shunting inhibitory cellular neural networks with chaotic external inputs, \textit{Chaos}, \textbf{23}, 023112.

\bibitem{Akh21} 
Akhmet, M. and Fen, M.O. (2014), Chaotification of impulsive systems by perturbations, \textit{Int. J. Bifurcation and Chaos}, \textbf{24}, 1450078.

\bibitem{Akh18} 
Akhmet, M.U. and Fen, M.O. (2014), Entrainment by chaos, \textit{Journal of Nonlinear Science}, \textbf{24}, 411-439.

\bibitem{Akh19} 
Akhmet, M. and Fen, M.O. (2014), Generation of cyclic/toroidal chaos by Hopfield neural networks, \textit{Neurocomputing}, \textbf{145}, 230-239.

\bibitem{Akh20} 
Akhmet, M.U. and Fen, M.O. (2015), Attraction of Li-Yorke chaos by retarded SICNNs, \textit{Neurocomputing}, \textbf{147}, 330-342.

\bibitem{Akh22} 
Akhmet, M. and Fen, M.O. (2016), \textit{Replication of Chaos in Neural Networks, Economics and Physics}, Springer-Verlag: Berlin, Heidelberg.

\bibitem{Brown93} 
Brown, R. and Chua, L. (1993), Dynamical synthesis of Poincar\'{e} maps, \textit{Int. J. Bifurcation and Chaos}, {\bf 3}, 1235-1267.

\bibitem{Brown96} 
Brown, R. and Chua, L. (1996), From almost periodic to chaotic: the fundamental map, \textit{Int. J. Bifurcation and Chaos}, {\bf 6}, 1111-1125.

\bibitem{Brown97} 
Brown, R. and Chua, L. (1997), Chaos: generating complexity from simplicity, \textit{Int. J. Bifurcation and Chaos}, {\bf 7}, 2427-2436.

\bibitem{Brown01} 
Brown, R., Berezdivin, R. and Chua, L. (2001), Chaos and complexity, \textit{Int. J. Bifurcation and Chaos}, {\bf 11}, 19-26.

\bibitem{Nemytskii60} 
Nemytskii, V.V. and Stepanov, V.V. (1960), \textit{Qualitative Theory of Differential Equations}, Princeton University Press: New Jersey.

\bibitem{Smale67} 
Smale, S. (1967), Differentiable dynamical systems, \textit{Bull. Amer. Math. Soc.}, \textbf{73}, 747-817.

\bibitem{Marotto78} 
Marotto, F.R. (1978), Snap-back repellers imply chaos in $\mathbb R^n$, \textit{J. Math. Anal. Appl.}, \textbf{63}, 199-223.

\bibitem{PLi07} 
Li, P., Li, Z., Halang, W.A. and  Chen, G. (2007), Li-Yorke chaos in a spatiotemporal chaotic system, \textit{Chaos, Solitons and Fractals}, \textbf{33}, 335-341.

\bibitem{Akin03} 
Akin E. and Kolyada, S. (2003), Li-Yorke sensitivity, \textit{Nonlinearity}, \textbf{16}, 1421-1433.

\bibitem{Kloeden06} 
Kloeden, P. and Li, Z. (2006), Li-Yorke chaos in higher dimensions: a review, \textit{Journal of Difference Equations and Applications}, \textbf{12}, 247-269.

\bibitem{Shi04} 
Shi Y. and Chen, G. (2004), Chaos of discrete dynamical systems in complete metric spaces, \textit{Chaos, Solitons \& Fractals}, \textbf{22}, 555-571. 

\bibitem{Shi05} 
Shi Y. and Chen, G. (2005), Discrete chaos in Banach spaces, \textit{Science in China, Ser. A: Mathematics}, \textbf{48}, 222-238. 

\bibitem{Izhikevich07} 
Izhikevich, E.M. (2007), \textit{Dynamical Systems in Neuroscience: The Geometry of Excitability and Bursting}, The MIT Press: Cambridge.

\bibitem{Pasemann2003} 
Pasemann, F., Hild, M. and Zahedi, K. (2003), SO(2)-networks as neural oscillators, \textit{Computational Methods in Neural Modeling Lecture Notes in Computer Science}, \textbf{2686}, 144-151.

\bibitem{Watanabe97} 
Watanabe, M., Aihara, K. and Kondo, S. (1997), Self-organization dynamics in chaotic neural networks, \textit{Control and Chaos, Mathematical Modelling}, \textbf{8}, 320-333.

\bibitem{Gon04} 
Gonz\'{a}les-Miranda, J.M. (2004), \textit{Synchronization and Control of Chaos}, Imperial College Press: London.

\bibitem{Hunt97} 
Hunt, B.R., Ott, E. and Yorke, J.A. (1997), Differentiable generalized synchronization of chaos, \textit{Phys. Rev. E}, \textbf{55}, 4029-4034.

\bibitem{Kapitaniak94} 
Kapitaniak, T. (1994), Synchronization of chaos using continuous control, \textit{Phys. Rev. E}, \textbf{50}, 1642-1644.

\bibitem{Kocarev96} 
Kocarev, L. and Parlitz, U. (1996), Generalized synchronization, predictability, and equivalence of unidirectionally coupled dynamical systems, \textit{Phys. Rev. Lett.}, \textbf{76}, 1816-1819.

\bibitem{Macau02} 
Macau, E.E.N., Grebogi, C. and Lai, Y.-C. (2002), Active synchronization in nonhyperbolic hyperchaotic systems, \textit{Phys. Rev. E}, \textbf{65}, 027202.

\bibitem{Pecora90} 
Pecora, L.M. and  Carroll, T.L. (1990), Synchronization in chaotic systems, \textit{Phys. Rev. Lett.}, \textbf{64}, 821-825.

\bibitem{Rulkov95} 
Rulkov, N.F., Sushchik, M.M., Tsimring, L.S. and  Abarbanel, H.D.I. (1995), Generalized synchronization of chaos in directionally coupled chaotic systems, \textit{Phys. Rev. E}, \textbf{51}, 980-994.

\bibitem{Abarbanel96}  
Abarbanel, H.D.I., Rulkov, N.F. and  Sushchik, M.M. (1996), Generalized synchronization of chaos: The auxiliary system approach, \textit{Phys. Rev. E}, \textbf{53}, 4528-4535.

\bibitem{Afraimovich01} 
Afraimovich, V., Chazottes, J.-R. and Cordonet, A. (2001), Nonsmooth functions in generalized synchronization of chaos, \textit{Phys. Lett. A}, \textbf{283}, 109-112.

\bibitem{Hale80} 
Hale, J.K. (1980), \textit{Ordinary Differential Equations}, Krieger Publishing Company: Malabar.

\bibitem{Franceschini80} 
Franceschini, V. (1980), A Feigenbaum sequence of bifurcations in the Lorenz model, \textit{J. Stat. Phys.}, \textbf{22}, 397-406. 

\bibitem{Sato83} 
S. Sato, M. Sano, and Y. Sawada,
\textit{Universal scaling property in bifurcation structure of Duffing's and of generalized Duffing's equations},
Phys. Rev. A, \textbf{28} (1983) 1654--1658.

\bibitem{Wiggins88} 
Wiggins, S. (1988), \textit{Global Bifurcations and Chaos}, Springer: New York.

\bibitem{Fink74} 
Fink, A.M. (1974), \textit{Almost Periodic Differential Equations}, Springer-Verlag: Heidelberg.

\bibitem{Hale91} 
Hale, J. and Ko\c{c}ak, H. (1991), \textit{Dynamics and Bifurcations}, Springer-Verlag: New York.

\bibitem{Sch99} 
Schuster, H.G. (1999), \textit{Handbook of Chaos Control}, Wiley-Vch: Weinheim.

\bibitem{Pyragas92} 
Pyragas, K. (1992), Continuous control of chaos by self-controlling feedback, \textit{Phys. Lett. A}, \textbf{170}, 421-428.

\bibitem{Pomeau80} 
Pomeau, Y. and Manneville, P. (1980), Intermittent transition to turbulence in dissipative dynamical systems, \textit{Commun. Math. Phys.}, \textbf{74}, 189-197.



\end{thebibliography}
\end{document}